\documentclass[twocolumn]{aastex62}

\usepackage{graphicx}

\graphicspath{{./}{figures/}}

\submitjournal{ApJ}

\shorttitle{t-SNE ETGs}
\shortauthors{Steinhardt et al.}

\begin{document} 

\title{A Method to Distinguish Quiescent and Dusty Star-forming Galaxies with Machine Learning}

\correspondingauthor{Charles L. Steinhardt}
\email{steinhardt@nbi.ku.dk}

\author[0000-0003-3780-6801]{Charles L. Steinhardt}
\affiliation{Cosmic Dawn Center (DAWN)}
\affiliation{Niels Bohr Institute \\
Lyngbyvej 2, 2100 K\o benhavn \O}

\author[0000-0003-1614-196X]{John R. Weaver}
\affiliation{Cosmic Dawn Center (DAWN)}
\affiliation{Niels Bohr Institute \\
Lyngbyvej 2, 2100 K\o benhavn \O}

\author{Jack Maxfield}
\affiliation{California Institute of Technology \\
1216 East California Boulevard, Pasadena, CA 91125}
\affiliation{Cosmic Dawn Center (DAWN)}
\affiliation{Niels Bohr Institute \\
Lyngbyvej 2, 2100 K\o benhavn \O}

\author[0000-0002-2951-7519]{Iary Davidzon}
\affiliation{California Institute of Technology \\
1216 East California Boulevard, Pasadena, CA 91125}
\affiliation{Infrared Processing and Analysis Center, 1216 East California Boulevard, Pasadena, CA 91125}

\author[0000-0002-9382-9832]{Andreas L. Faisst}
\affiliation{California Institute of Technology \\
1216 East California Boulevard, Pasadena, CA 91125}
\affiliation{Infrared Processing and Analysis Center, 1216 East California Boulevard, Pasadena, CA 91125}

\author{Dan Masters}
\affiliation{California Institute of Technology \\
1216 East California Boulevard, Pasadena, CA 91125}
\affiliation{Jet Propulsion Laboratory, 4800 Oak Grove Dr, Pasadena, CA 91109}

\author{Madeline Schemel}
\affiliation{California Institute of Technology \\
1216 East California Boulevard, Pasadena, CA 91125}
\affiliation{Niels Bohr Institute \\
Lyngbyvej 2, 2100 K\o benhavn \O}

\author{Sune Toft}
\affiliation{Cosmic Dawn Center (DAWN)}
\affiliation{Niels Bohr Institute \\
Lyngbyvej 2, 2100 K\o benhavn \O}

\begin{abstract}
Large photometric surveys provide a rich source of observations of quiescent galaxies, including a surprisingly large population at $z>1$.  However, identifying large, but clean, samples of quiescent galaxies has proven difficult because of their near-degeneracy with interlopers such as dusty, star-forming galaxies.  We describe a new technique for selecting quiescent galaxies based upon t-distributed stochastic neighbor embedding (t-SNE), an unsupervised machine learning algorithm for dimensionality reduction.  This t-SNE selection provides an improvement both over UVJ, removing interlopers which otherwise would pass color selection, and over photometric template fitting, more strongly towards high redshift.  Due to the similarity between the colors of high- and low-redshift quiescent galaxies, under our assumptions t-SNE outperforms template fitting in 63\% of trials at redshifts where a large training sample already exists.  It also may be able to select quiescent galaxies more efficiently at higher redshifts than the training sample.
\end{abstract}

\keywords{}


\section{Introduction} 
\label{sec:intro}

Although not highly celebrated, perhaps the most influential discovery in the history of modern astronomy has been gradually finding that galaxies do not take on arbitrary properties, spanning the entire range of theoretically possible spectra.  As a result, it has been possible to produce meaningful surveys of faint galaxies using photometry, with only a very limited amount of information about the full spectral energy distribution (SED).  Most commonly, a series of templates \citep{Bruzual2003,Maraston2009,Brown2014} are fit to photometric colors with one of several competing techniques (cf. \citealt{LePhare,Brammer2008,Kriek2009}) in order to produce a best-fit set of parameters.

Fundamentally, the goal of photometry is to map observed colors to galaxy properties.  The validity of this technique therefore requires two additional assumptions.  First, the mapping between observed colors and galaxy properties must be surjective, i.e., any specific combination of colors must only be produced by one set of galaxy properties.  Otherwise, the colors are insufficient to break degeneracies between different possible galaxy models.  Second, due to the complexity of calculating synthetic templates, current codes use a precompiled library of discrete models.  Therefore, it is also necessary to assume that similar colors map to similar properties, to the point that it is possible to interpolate between nearby points with a known mapping.

Interpolation presents a considerable challenge, because there are often $\sim 10$ galaxy parameters which one would like to fit, and this produces too large of a search space.  Fortunately, we have discovered a series of scaling and other relations between observed galaxy parameters, including the `fundamental plane'~\citep{Gudehus1973,Pahre1998,Bernardi2003} between radius, velocity dispersion, and surface brightness, the `star-forming main sequence'~\citep{Brinchmann2004,Noeske2007,Peng2010,Speagle2014}, and a similar sequence for quasar accretion~\citep{Steinhardt2010a,Steinhardt2011}.  Because galaxies do not span the entirety of this $\sim 10$-dimensional space, it is natural to consider first mapping to a smaller space which can be entirely searched, then running similar algorithms.  

Previous work has shown that dimensionality reduction via a self-organizing map (SOM; \citealt{Kohonen1982}) can be used to map photometry to a two-dimensional space suitable for redshift determination \citep{Masters2015,Hemmati2019}.  The SOM spreads objects out approximately equally, dedicating more cells to more common types of objects.  In this work, we use a related technique, t-Distributed Stochastic Neighbor Embedding (t-SNE;  \citealt{vanderMaaten2008,vanderMaaten2014}), which similarly produces a map with reduced dimension, but will produce a sparser mapping in an attempt to preserve structure and relative distance. 

It might be hoped that combining such a map with observed spectroscopic redshifts will provide the basis for unsupervised machine learning-derived photometric redshifts.  In practice, redshifts determined by the SOM may produce a lower bias, suitable for several applications to Euclid \citep{Massey2013,Masters2015,Masters2017,Masters2019,Hemmati2019} and other upcoming surveys.  For relatively common objects where high-quality training data is available, it is also possible to directly use machine learning to model other galaxy parameters \citep{Krakowski2016,Siudek2018,Davidzon2019}.  However, at present photometric redshifts derived from template fitting remain competitive with those from unsupervised machine learning and for exotic outliers with few counterparts in a training sample are typically superior \citep{Masters2015,Hemmati2019}.  In effect, at this point there is more information in the theoretical templates derived from current models than there is from observed spectroscopic redshifts for rare but well-understood objects.

However, it has been possible to apply machine learning techniques to a wide variety of astronomical problems for which observations indeed provide more information than theoretical models.  Recent work has included the use of t-SNE to derive stellar chemical abundances \citep{Anders2018} and spectral information \citep{Traven2017}, as well as the use of Convolutional Neural Networks to measure galaxy morphology \citep{Dieleman2015, DominguezSanchez2018, Cheng2019, Hausen2019} and shape \citep{Ribli2019}, perform light profile fitting \citep{Tuccillo2018}, identify mergers \citep{Bottrell2019}, estimate cluster masses \citep{Ho2019}, and classify supernovae \citep{Muthukrishna2019}.

For the same reason, unsupervised machine learning should be expected to perform better than template fitting for objects which are poorly modeled by current theory.  In this work, we use a combination of t-SNE and current observations to develop a new selection for high-redshift, quiescent galaxies.  In \S~\ref{sec:estimators}, the underlying assumptions and a more formal definition of a quiescent estimator is given.  The new estimator is described in \S~\ref{sec:tsne}.  We then evaluate the success of this algorithm in \S~\ref{sec:comparison}.  

\section{Quiescence Estimators and Varying Assumptions}
\label{sec:estimators}

In this work, we develop a method based upon the unsupervised machine learning algorithm t-SNE to select quiescent galaxies from photometric surveys.  Let $\left\{\vec{x_i} \in \mathcal{R}^k\right\}$ be the set of photometric measurements provided by the survey, where each individual $\vec{x_i}$ has $k$ components, representing one object in $k$ bands.  Each specific galaxy might be quiescent; denote this by $q_i \in \{0,1\}$.  A method of quiescent galaxy detection consists of an estimator\footnote{Note that this is a more general definition than required for previous estimators, which can be given an individual $\vec{x_i}$ and produce $Q(\vec{x_i}) = q_i$.  The machine learning classification developed here can only operate on the entire set $\{\vec{x_i}\}$ simultaneously, and is meaningless for individual objects.} $Q(\{\vec{x_i}\})$ for $\{q_i\}$.  Although the true $q_i \in \{0,1\}$, some estimators instead return a probability of quiescence $q_i \in [0,1]$.

Current estimators use information learned from past analysis of galaxies to produce a static mapping $Q(\vec{x_i}) \rightarrow q_i$.  The most commonly used example is a color-color selection such as a UVJ diagram \citep{Strateva2001,Baldry2004,Wuyts2007,Williams2009,Muzzin2013,vanderWel2014,Leja2019}, which is a mapping from only three of the $k$ bands (chosen or adjusted to approximate U, V, and J bands in rest-frame colors) to a quiescence estimator, selecting a region of ratios between adjacent bands which is populated primarily by quiescent galaxies.  A far more complex mapping is produced by photometric template fitting \citep{LePhare,Brammer2008,Kriek2009}, which produces spectra for various combinations of model parameters and uses them to find a best-fit spectrum for each $\vec{x_i}$.  The model parameters corresponding to that spectrum determine the specific star formation rate (sSFR, or SFR per unit stellar mass), and applying a threshold to the sSFR produces a quiescent estimator.  Both methods assume that our knowledge of {\em typical} galaxies is sufficient to produce a mapping $Q$ that will be valid for {\em all} galaxies, and in the case of template fitting that knowledge includes a mapping from model parameters to spectrum through stellar population synthesis (cf. \citealt{Conroy2009,Conroy2013}).  

The unsupervised machine learning method developed here does not require stellar population synthesis or any other astrophysics, but rather attempts to let the data itself determine the proper quiescence estimators.  Conceptually, the success of such an algorithm relies upon three key assumptions:
\begin{enumerate}
    \item {There is a surjection $Q(\vec{x_i}) \rightarrow q_i$ from photometric measurements in the survey bands to quiescence.  Unlike the methods described above, {\em it is not necessary to know any properties of that surjection}, but merely that one exists.}
    \item {If $x_1$ and $x_2$ are nearby in the $k$-dimensional photometric color space, it is very likely that $q_1 = q_2$.}
    \item {There is a mapping $T:\mathcal{R}^k \rightarrow \mathcal{R}^2$ from the $k$-dimensional vector space to a lower-dimensional space, in this case two-dimensional, in which the previous two properties continue to hold.}

\end{enumerate}

The first assumption is straightforward, as it is minimum necessary foundation for photometry as a valid astronomical technique.  In regions where the second assumption does not hold, it means that quiescent and non-quiescent galaxies will be very nearly degenerate.  One example would be the near-degeneracy between age and extinction in photometric template fitting \citep{Gallazzi2005}.  Thus, photometry will be insufficient to determine quiescence with high certainty for such galaxies regardless of the methodology employed.

The third assumption is necessary because unsupervised machine learning requires a training set, and galaxy photometry is sparse in $\mathcal{R}^k$.  Thus, we first map galaxies $T:\left\{\vec{x_i}\right\}\rightarrow\left\{\vec{x^{\prime}_i}\right\}$ into $\mathcal{R}^2$, then find a surjection $Q(\left\{\vec{x_i}\right\}) \rightarrow \left\{q_i\right\}$ that produces a quiescence estimator from a reduced space in which individual galaxies are likely to have many close neighbors.  

If all of these assumptions hold, it is then possible to produce an estimator for whether any specific galaxy should be classified as quiescent by looking at neighboring galaxies for which $q$ has been well measured and letting those neighbors vote.  We evaluate the correctness of this estimator in \S~\ref{sec:comparison}, finding that on average it is more successful than template fitting, more strongly so towards high redshift.  However, there is ultimately value in using both approaches, one which is based upon astrophysical knowledge about the physics of galaxy evolution and another which is entirely ignorant of that physics and only given examples of quiescent and non-quiescent galaxies, letting the data alone predict quiescence.

\section{Using t-SNE to Select Quiescent Galaxies}
\label{sec:tsne}

The most successful existing methods for  photometric quiescent galaxy selection are variations on color-color selection, in which galaxies are mapped into a two-dimensional space based upon the two slopes described by a set of three specific rest-frame photometric bands, then a region is identified which is populated primarily by quiescent galaxies \citep{Williams2009}.  Every galaxy within that region is selected as quiescent, and the remainder as selected as non-quiescent.

The machine learning method here, although is it constructed from a very different toolkit, is essentially an improved version of this familiar color-color selection.  It similarly finds regions populated primarily by quiescent galaxies, but using all available bands, so that all available information can be used in the selection.  Further, quiescent galaxies are selected from any of a potentially large number of tiny regions.  As a result, with a sufficiently high-quality training sample, it becomes possible to exclude dusty star-forming galaxies which are {\em nearly} identical, but not entirely identical, to quiescent galaxies and avoid selecting them.  Naturally, if star-forming galaxies are truly degenerate with quiescent galaxies, no algorithm can distinguish them, in which case the first assumption (\S~\ref{sec:estimators}) would be violated because $Q(\vec{x_i}) \rightarrow q_i$ would no longer be surjective.  Similarly, if in practice measurement uncertainties are large enough to partially or fully restore degeneracies, it will again become impossible to select all quiescent galaxies but no interlopers.

\begin{figure*}[ht!]
\plotone{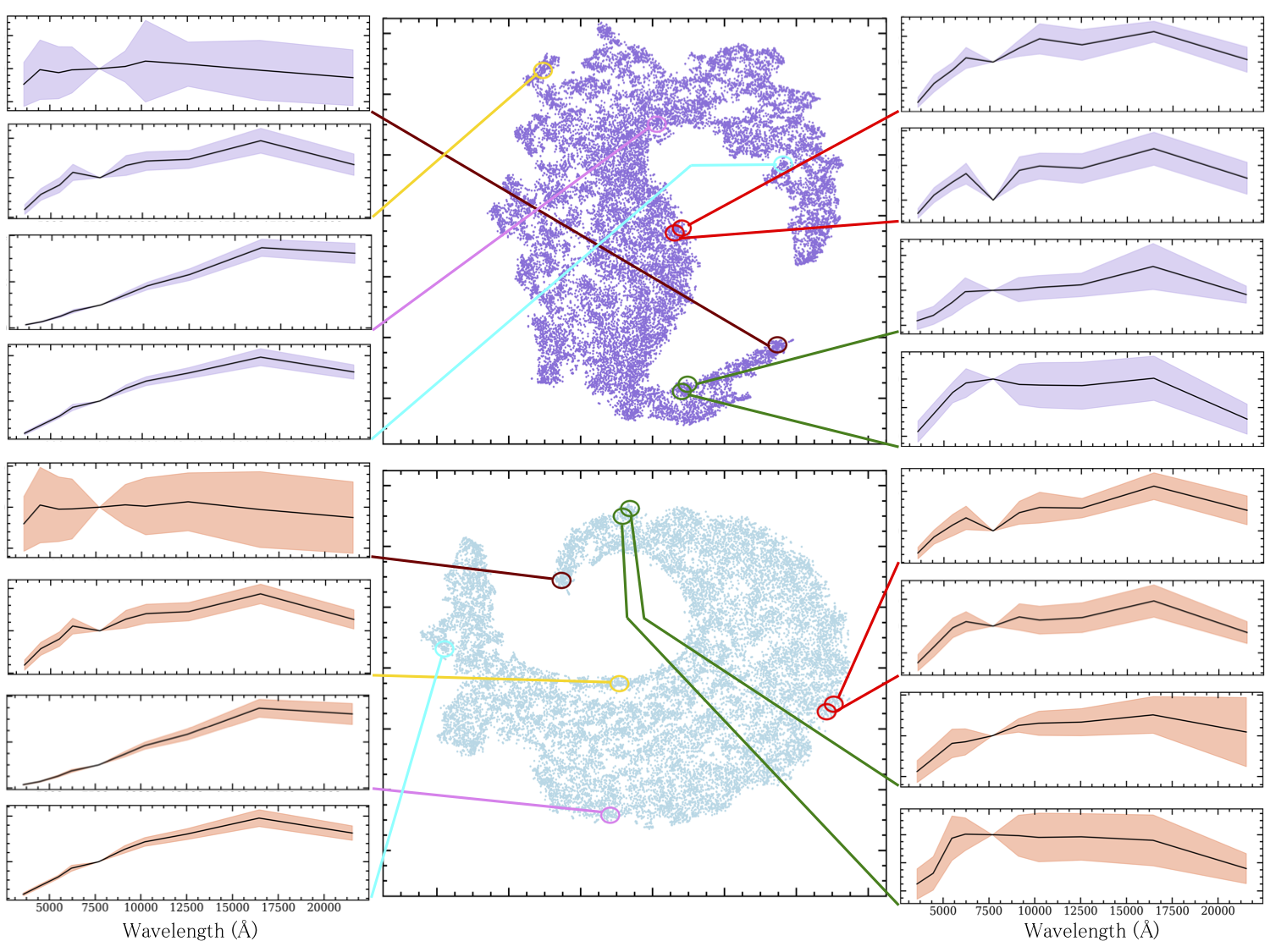}
\caption{(a) t-SNE map of all galaxies at $1.0 < z < 1.1$ in the ULTRAVISTA catalog \citep{McCracken2012} for which MIPS coverage is available.  A narrow redshift range is necessary, since otherwise the primary structure shown in the map would indicate redshift.  Galaxies with similar SEDs end up in neighboring locations, and galaxies with dissimilar SEDs end up far apart.  Mean SEDs with 1-$\sigma$ envelopes are calculated from the black circles indicated. (b) A second t-SNE map for the same catalog, produced with a different random seed and initial ordering of the sample.  The eight sample SEDs shown are the same for both maps, with corresponding colors identifying the same SEDs. In each case, the range of SEDs within the indicated circle is shown, with all SEDs normalized to a common z-band flux.}\label{fig:tsne}
\end{figure*}

The algorithm we select, t-SNE \citep{vanderMaaten2008,vanderMaaten2014}, is an unsupervised machine learning algorithm for dimensionality reduction designed for the visualization of high-dimensional datasets.  We first use t-SNE to produce a map $T$ in which galaxies with similar photometric SEDs are placed in nearby locations, while galaxies with dissimilar photometric SEDs are further away (Fig. \ref{fig:tsne}a).  For the figures shown in this work, maps were constructed in rest-frame magnitude space, and different features of the dataset would be revealed using different units or distance metrics.  

Note that the two coordinates produced by t-SNE are arbitrary and do not represent any sort of basis for the space. A galaxy further in the x-direction has not become more `x-like', but is merely dissimilar from galaxies further to the left.  Further, t-SNE is a randomized algorithm, and running it on the same dataset with different initial conditions will produce the same topology but a different map (Fig. \ref{fig:tsne}b).  

\begin{figure*}[ht!]
\plotone{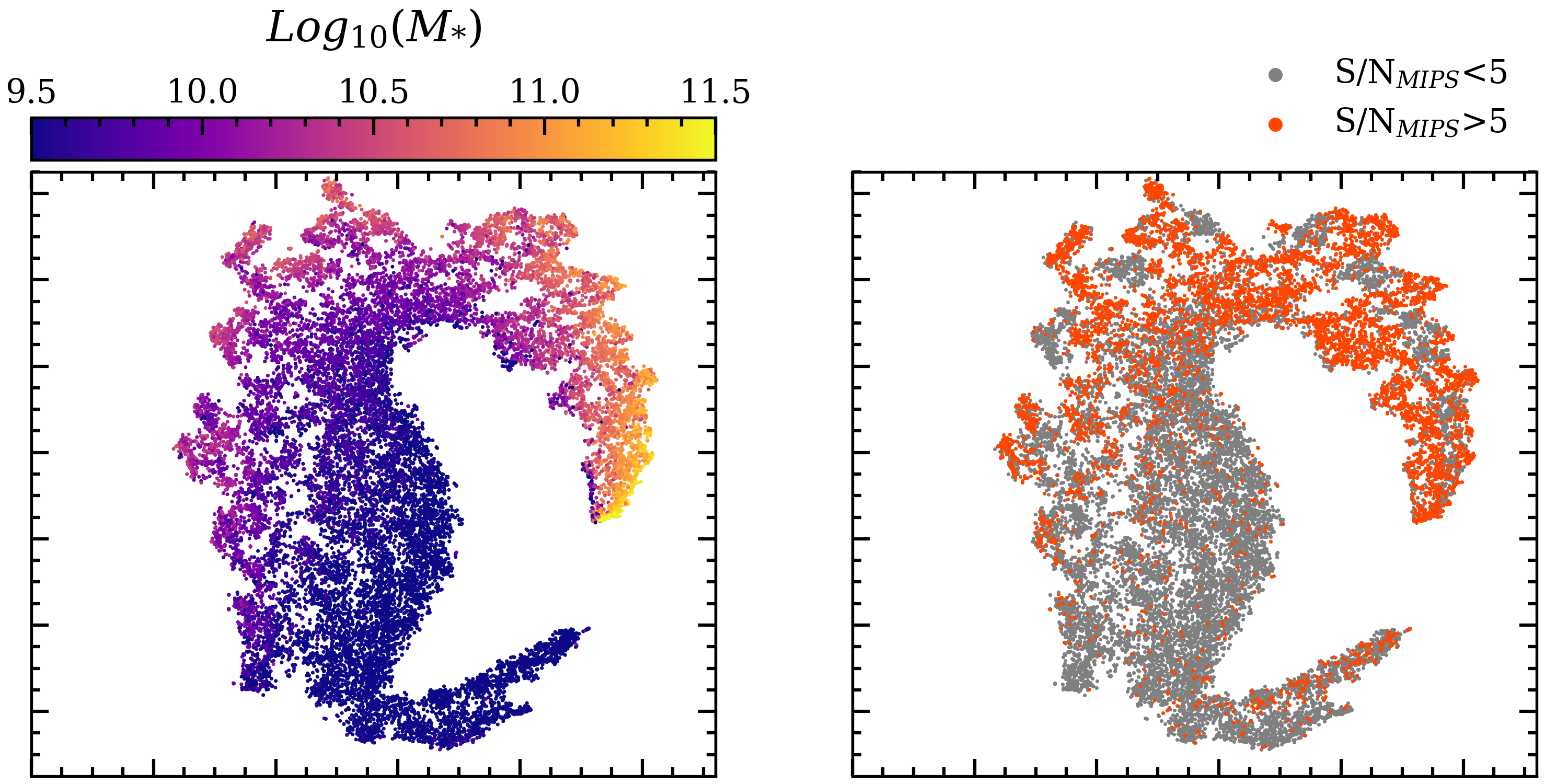}
\caption{(left) Best-fit stellar mass for the galaxies in the ULTRAVISTA sample from Fig. \ref{fig:tsne}.  Galaxies with similar best-fit stellar masses cluster together.  Thus, it is possible to predict the photometric template fitting-determined stellar mass of a galaxy without actually running those template fitting codes on the object in question by instead looking at the masses of its neighbors.  The same is true of many other galaxy properties which currently require template fitting.  (right) Summary of MIPS detections for the low-z sample. Sources with credible detections (S/N\textgreater 5; orange) are clustered with respect to those without credible detections (S/N\textless 5; grey).  \label{fig:properties}}
\end{figure*}

To this point, the t-SNE map has been produced without any direct information about galaxy quiescence or any other astronomical properties.  However, because photometry is an indicator of astrophysical quantities, the map resulting from arranging galaxies based upon their photometry has also arranged them by these quantities.  It is therefore possible to predict the properties such as stellar mass that would be found by photometric template fitting without the need to run template fitting codes, merely by looking at the results of running those codes on nearby galaxies (Fig. \ref{fig:properties}, left).

Most importantly for our purposes, the same is very likely true for quiescence.  It is certainly true that a t-SNE map can predict whether template fitting will determine that a galaxy is quiescent, but this has limited utility.  After all, the primary advantage of unsupervised machine learning is avoiding the need to make the assumptions required to produce templates.  Instead, we train our predictor on the most successful technique applied to the ULTRAVISTA catalog, combining a successful (rest-frame) UVJ color selection with a non-detection in the Spitzer Space Telescope MIPS 24$\mu$m band \citep{MIPS}, an independent indicator of hot dust and likely also a high star-formation rate \citep{Rieke2009}.

Galaxies with MIPS detections lie primarily within a set of contiguous regions on the t-SNE map (Fig. \ref{fig:properties}).  Further, nearly every galaxy within that region has a MIPS detection, while nearly every galaxy lying elsewhere does not.  Similarly, galaxies classified as quiescent (UVJ selected but no MIPS detection) lie within distinct regions (Fig. \ref{fig:tsne_map}).  Therefore, it is natural to produce an estimator that examines the objects within a small neighborhood on the map with known classification as quiescent or non-quiescent and lets them vote on whether a new object is likely to be quiescent and on the confidence in that prediction.  If the number of such neighborhoods were known {\em a priori}, that information could be used to produce a further improved predictor (cf. \citealt{Turner2019}).  However, avoiding the imposition of this condition allows t-SNE to attempt to detect of all types of quiescent galaxies and all types of dusty star-forming galaxies without imposing any prior expectations on how many distinct types exist.

\subsection{Information Used for t-SNE Mapping}
The \citet{Laigle2016} catalog includes over 30 bands from NUV out NIR , some which overlap with alternative bands at similar wavelength. However due to differences in on-sky coverage, many of those bands are only available for a fraction of the catalog.  Because t-SNE relies on Euclidean distances, which cannot be calculated for vectors of different dimension, we restrict our analysis to $u, B, r, i^{+}, z^{++}, Y, J, H, K_{s}$ bands, which are available as statistically-significant detections for most of the catalog. As described in Table 1 of \citet{Laigle2016}, their 3$\sigma$ depths vary across the field and range from 23.4 ($K_{s}$) to 27.0 ($B$). Due to completeness, a mass cut is made between $8.5 < \log_{10}(M_{*}) < 11.5$. Based on photometric redshift solutions, we isolate a low redshift sample within $0.9 < z < 1.1$ containing 19,774 galaxies (17\% quiescent) and a high redshift sample within $1.9 < z < 2.1$ containing 7,524 galaxies (6\% quiescent)  When training and test samples are made, in every case objects are randomly selected from within the relevant redshift boundaries.

\subsection{Definition of ``True" Quiescent Galaxies}
\label{subsec:truedef}
What any learning algorithm attempts to do is predict how new data would have been labeled if it were part of that training sample.  A significant issue in training t-SNE selection is therefore that the labeling, in the ideal case, will ultimately be exactly as good as the training sample.  Thus, using t-SNE to select quiescent galaxies relies on a good selection of the ground truth of quiescence for the training sample.

Ideally, this would be done spectroscopically, using specific lines as tracers of SFR and stellar mass.  However, spectroscopy is available for only about 1\% of the \citet{Laigle2016} catalog.  Further, objects with spectroscopic followup have often been targeted because of specific photometric properties, so that a complete spectroscopic sample would be even smaller.  It is therefore necessary to rely on photometry to select ``true'' quiescent galaxies in the training and test samples.  Several methods have been proposed for using photometry to select quiescent galaxies, including the new method developed here.  Of these, each has known flaws:

\begin{itemize}
    \item {Static color selection, such as UVJ, provides a fairly good proxy for quiescence while minimizing contamination.  However, some objects with very high SFR (and MIPS flux) will  pass the UVJ selection (Fig. \ref{fig:uvjrocc}a and related discussion).  Quiescent galaxies can also lie outside the UVJ region \citep{DominguezSanchez2016}. A possible solution to these limitations is replacing the rest frame $U$ band with near-UV (NUV) and increase the color leverage by using $NUV-r$ vs.\ $r-K$  \citep[or $r-J$, see respectively][]{Arnouts2013,Ilbert2013}. }
    \item{MIPS is only successful at selecting the galaxies which are brightest at 24$\mu$m (COSMOS does not have uniform MIPS coverage, so the detection threshold varies significantly). Therefore, many star-forming galaxies, including ones which at low mass but high redshift lie \emph{above} the main sequence, will not be detected by MIPS.  
    
     At $z \sim 2$, this potential mischaracterization of lower-SFR (likely lower mass) star-forming galaxies will be significantly worse than at $z \sim 1$.  Thus, there is a strong redshift dependence in this definition of a true quiescent galaxy.  This effect will therefore underestimate the quality of a selection trained at one redshift but tested at another (\S~\ref{subsec:test12}).}
     
    \item{Photometric template fitting and the resulting sSFR attempts to calculate a quantity which can be most directly interpreted as ``true" quiescence.  However, photometric template fitting is also known to produce significant errors in SFR (cf. \citealt{Laigle2019}), and therefore even less reliable sSFR, dividing that by an estimated stellar mass.  Often there is insufficient multi-wavelength coverage to use any other method (except for UVJ or some other two color selection), and photometric template fitting is used by default.  However, as shown in \S~\ref{sec:comparison}, these flaws in using template fitting to estimate SFR mean that best-fit sSFR is not a particularly good proxy for quiescence, even though true sSFR would be.}
\end{itemize}

Given the available options, in this work a combination of UVJ and MIPS selection is used to define ground truth for quiescence.  When a sufficiently large spectroscopic sample can be produced, it would be ideal to then recalibrate the t-SNE predictor based upon this improved definition of true quiescence.

\subsection{Training and Test Samples}

Since t-SNE has no knowledge of astronomy, it must be provided with a training set consisting of identified quiescent and non-quiescent galaxies in order to produce a predictor.  Unlike algorithms such as a self-organizing map \citep{Kohonen1982,Masters2015}, t-SNE does not produce a static transformation from the higher-dimensional space to the lower-dimensional one, but rather produces a mapping that extremizes a global penalty score for the entire sample.  Thus, adding a new object requires recalculating the entire t-SNE map, and may alter the positions of every object.  This means that t-SNE is a poor choice for real-time analysis, because it is not possible to precompute a static surjection $Q(\vec{x})$.  However, if the training sample is already large compared with the test sample, the entire test sample can be processed in approximately the same time as one object.

\begin{figure*}[ht!]
\plotone{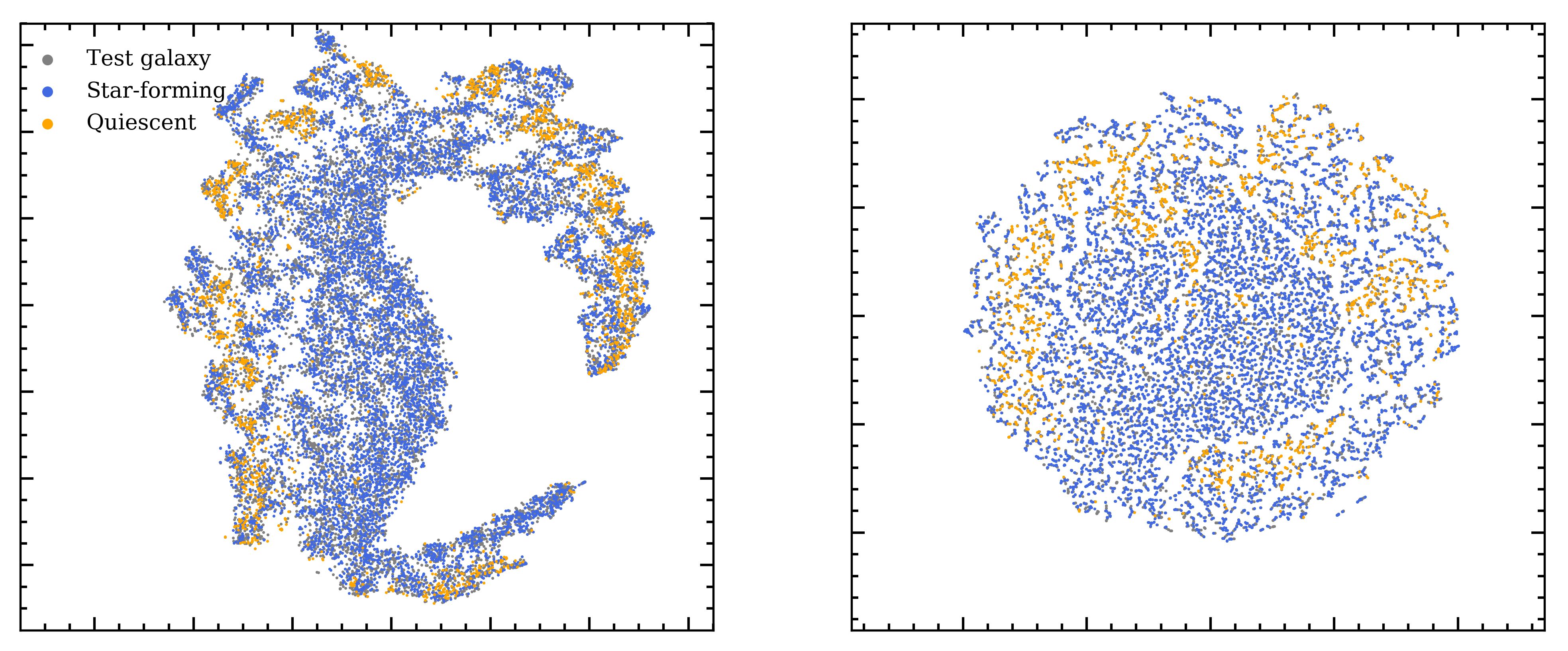}
\caption{(left) Mapping produced by t-SNE for a combination of training (blue labeled star-forming and orange labeled quiescent) and test (gray) sets, both taken from z$\sim$1. The quiescent galaxies cluster within the map, implying that galaxies in the test sample should be labeled as star-forming or quiescent in a similar manner.  (right) A second mapping produced with a smaller value of perplexity, which effectively optimizes the map for finding more local rather than global structures.  Different types of structure will be produced by different choice of t-SNE hyperparameters, so using t-SNE for predicting astronomical properties requires careful tuning.  Maps are produced with perplexity hyperparameters of 30 (left) and 7 (right).
\label{fig:tsne_map}}
\end{figure*}

Therefore, the estimator developed here first uses t-SNE to arrange the union of both training and test samples.  Galaxies described by their rest-frame photometry in the higher dimension space are mapped by t-SNE with a perplexity of 30 over 1000 iterations. Once converged, labels are applied to the training set to denote quiescence (Fig.~\ref{fig:tsne_map}).  It should be noted that the choice of perplexity and other settings (conventionally called hyperparameters in order to distinguish them from parameters, which instead belong to the model) makes a substantial difference and is part of the t-SNE tuning process.  Perplexity formally is defined in terms of the Shannon entropy of the system \citep{vanderMaaten2008}, so that higher values produce a wider search which results in more strongly weighting global structure and lower values similarly reveal more local structure.  The choice of perplexity is effectively a prediction of the number of neighbors which should be used in determining the properties of a galaxy, and therefore depends not only on the underlying distribution of galaxy properties but also on sample size and selection. 

Because the quiescent galaxies cluster within the training sample, if galaxies in the test sample are drawn from a very similar population, they should have the same label as their neighbors (Fig.~\ref{fig:tsne_map}).  This is also a corollary of the three assumptions listed in \S~\ref{sec:estimators}.  Objects in the test sample are therefore classified as quiescent when the quiescent fraction of $m$ neighboring training galaxies is $f_Q > f_{\textrm{min}}$.   

A natural choice of $f_{\textrm{min}}$ might seem to be 0.5, letting a majority of nearby objects determines the label.  However, in practice the optimal choice depends upon how many objects are scattered via measurement errors to incorrectly be close neighbors.  Therefore, the optimal choice of $f_{\textrm{min}}$ depends upon a combination of uncertainties and the underlying true fraction of quiescent galaxies in the training sample.  The choice of $f_{\textrm{min}}$ also depends upon the desired relationship between quantity and quality in the resulting catalog; a higher value of $f_{\textrm{min}}$ will result in fewer false positives but more false negatives (e.g., Fig. \ref{fig:roc11}).  It is common to see this tradeoff referred to in machine learning literature as one between precision (true positives / total positives), a measure of sample quality, and either recall or sensitivity, two terms referring to the true positive rate, which is a measure of sample quantity.

\section{Comparison with Template Fitting}
\label{sec:comparison}

We construct a series of tests to compare the effectiveness of t-SNE in selecting ``quiescent'' galaxies (defined here as UVJ selected but not MIPS detected) with that of photometric template fitting.  Since our fiducial definition of a quiescent galaxy includes a two-color (UVJ) selection, it is difficult to test t-SNE against two-color selection.  However, in \S\ref{subsec:testuvj}, we evaluate whether t-SNE is more likely to discard dusty star-forming galaxies with MIPS detections than a standard two-color selection.

In all cases, the comparison is done on the ULTRAVISTA photometric catalog \citep{McCracken2012}, with data drawn from the \citet{Laigle2016} catalog providing improved reductions and additional ancillary data.  Currently, the most reliable method for differentiating between dusty, star-forming interlopers and bona fide quiescent galaxies in ULTRAVISTA requires additional observations, using Spitzer/MIPS to search for 24$\mu$m emission characteristic of hot dust, and therefore a dusty star-forming galaxy.  A successful classification is therefore defined as predicting correctly from $u, B, r, i^{+}, z^{++}, Y, J, H, K_{s}$ bands whether a galaxy will both be selected by (rest-frame) UVJ and have no discernible MIPS 24$\mu$m detection with a S/N$>$5, at which threshold it would instead be considered a dusty star-forming galaxy.

The sample used is described in detail in \cite{Laigle2016}, which provides robust multi-wavelength rest-frame magnitudes.  Galaxies are then cross-matched with a FIR/mm catalog \citep{Sanders2007,Jin2018} over the same footprint.  Objects with MIPS $\mathrm{ SNR} < 3$ are considered too faint for a MIPS non-detection to be a reliable indicator of quiescence, and are not included in the sample.  Although color selection and template fitting are static methods, t-SNE requires a training sample.  Therefore, for t-SNE the catalog must be divided into a training sample and test sample.

Template fitting and t-SNE selection are compared in their ability to solve two different problems.  First, to examine the case of a well-understood domain, the $0.9 < z < 1.1$ population is divided up into equal-sized, disjoint training and test samples.  t-SNE is given the entire training sample (in principle, so are other methods, but they do not change based upon new information) and a list of which training objects have been identified as quiescent.  The methods are also given the entire test sample and its NIR photometry, but no information about which test objects have MIPS detections.  Each method produces a catalog of test objects classified as quiescent, and is evaluated on both false positive and false negative rates.  

Second, template fitting and t-SNE selection are also evaluated on their ability to determine which galaxies are quiescent in an unexplored domain.  The training sample consists of the entire $0.9 < z < 1.1$ catalog, but the test sample consists of the $1.9 < z <2.1$ catalog.  Because the $z\sim2$ galaxy population is not identical to that at $z\sim1$ (cf. \citealt{Speagle2014} for star-forming galaxies and \citealt{vanderWel2014} for quiescent ones), this presents a far more difficult problem for machine learning, which has no knowledge of astronomy or any expected redshift evolution.   Methods are given rest-frame colors from the \citet{Laigle2016} catalog, so any errors in photometric redshift determination for the test sample will result in all three methods making predictions from incorrect inputs at the same rate.  Errors in photometric redshift determination for the training sample will degrade the efficiency of t-SNE, but not the other methods.

\subsection{Comparison of Estimators in a Well-Explored Domain}
\label{subsec:test11}

We first consider these estimators in a domain which is already well-explored.  Both the training and test samples are drawn from the same catalog at the same photometric redshift of $0.9 < z_\textrm{phot} < 1.1$, with ``true'' quiescent galaxies defined as in \S~\ref{subsec:truedef}. A typical use case might be producing a catalog of quiescent galaxies for a large photometric survey with limited spectroscopy.  In that case, the additional bands or spectroscopic followup sufficient to produce confirmed quiescent galaxies would only exist for a small fraction of the full catalog, but could provide a high-quality training sample.

\begin{figure*}[ht!]
\plotone{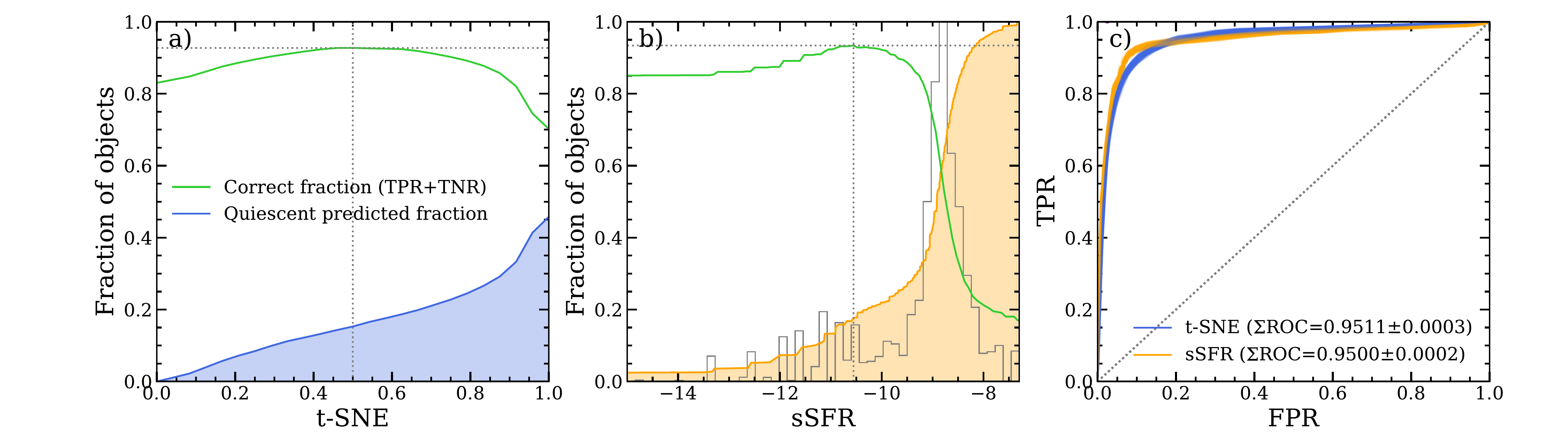}
\caption{(a-b) Correct (sum of of true positive and true negative) prediction fractions for (a) t-SNE and (b) photometric template fitting (green) as a function of the threshold used for both training and test sample drawn from the same $0.9 < z_\textrm{phot} < 1.1$ ULTRAVISTA catalog.  The sSFR distribution (black, panel b) and cumulative distribution (blue/orange in panels a/b) are also shown, as well as the maximum TPR+TNR achieved and their corresponding thresholds (grey dotted lines). (c) Receiver operating characteristic (ROC) curve for both methods over 10,000 random draws of training and test samples.  t-SNE (blue) outperforms photometric template fitting (orange) in 63\% of trials.}  With different tuning, t-SNE selection could constructed to outperform sSFR selection either for high-quantity samples (illustrated here) or instead for high-quality samples, but not for both use cases simultaneously. A typical threshold used to identify quiescent galaxies based on log$_{10}$(sSFR) is -10. \label{fig:roc11}
\end{figure*}

Color selection has no free parameters, and therefore is completely defined and produces a fixed error rate, both for false positives and false negatives.  The other two methods do have tunable parameters.  For photometric template fitting, a large number of selections (choice of templates, grid spacing, other fit parameters and hints) were made in the ULTRAVISTA catalog used (described in detail in \citealt{Laigle2016}), and cannot be altered for this test.  However, the choice of sSFR threshold used to determine quiescence is an additional parameter, and a higher threshold will reduce both true and false positives for quiescence (Fig. \ref{fig:roc11}b).

For t-SNE, there are similarly several hyperparameters required to produce an estimator.  The most significant for producing a map is perplexity, which governs the relative importance of local neighbors compared with more distant ones.  Once the map is produced, the definition of close neighbors, how many training sample neighbors are chosen, and threshold fraction of quiescent training sample neighbors are additional choices.  These choices must be made differently for every specific use case, because two identical galaxies will end up at different distances on the t-SNE map depending upon properties of other galaxies, sample selection, sample size, perplexity, and t-SNE grid size.  For this specific test, we experimentally determined that perplexity 30 maximized $\Sigma$ROC (defined below) for the figures shown.  Then, setting the threshold for the required fraction of quiescent neighbors to label a galaxy as quiescent produces a similar tradeoff between true and false positive rates as for template fitting (Fig. \ref{fig:roc11}a).   The total accuracy shown for each estimator is a combination of the true quiescent galaxy fraction and a false positive fraction, (TP+TN)/total.  The same accuracy could be produced from, e.g., more true quiescent galaxies with more false positives or fewer of each.  Thus, the maximum accuracy will lie at the cutoff  where a marginal change in cutoff will result in an equal change in both true and false positives.

The appropriate tool for comparing these estimators is a receiver operating characteristic (ROC) curve, a tool which is common for assessing the quality of medical diagnostics producing a boolean answer \citep{Albeck1990,Baker2003,Fawcett2006}.  A random estimator containing no information can be produced lying anywhere along the dashed diagonal, e.g., randomly selecting 40\% of galaxies to be quiescent will result in a 40\% of quiescent galaxies selected as quiescent (true positive) as well as 40\% of star-forming galaxies selected as quiescent (false positive).  The best estimators have ROC curves lying as close as possible to the top left, corresponding to 100\% true positives with no false positives.  

We find that the t-SNE ROC curve is comparable to the template fitting ROC curve (Fig. \ref{fig:roc11}c), with a slightly different shape.  For some desired true positive rates, a t-SNE method will produce a corresponding sample with fewer false positives, but for some true positive rates, template fitting is more successful.  With different hyperparameters, t-SNE selection could constructed to outperform sSFR selection either for any specific true positive rate desired (illustrated for high-quantity samples in Fig. \ref{fig:roc11}c) but not for all true positive rates simultaneously and with a lower overall success rate. 

A related statistic\footnote{In previous literature in medical diagnostics and later machine learning, this is described as "area under the curve", or AUC, rather than in terms of the integral.  Since astronomers should be more comfortable with a description in terms of calculus, we choose to do so instead.} is $\Sigma\textrm{ROC} \equiv \int_0^1\textrm{ROC(TPR)}d\textrm{FPR}$, commonly used in machine learning to consider the quality of estimators across all possible thresholds.  Prior to selecting a threshold, both template fitting (using sSFR) and t-SNE (using fraction of quiescent neighbors) in a ranked ordering of the entire test sample by likelihood that the object is quiescent.  The selection of a threshold then divides the sample into two groups, labeling the more likely group with $q=1$ and the remainder as $q=0$.  The $\Sigma\textrm{ROC}$ corresponds to the probability that a randomly selected quiescent galaxy is ranked higher than a randomly-selected star-forming galaxy (cf. \citep{Bradley1997}), and produces a similar result the Wilcoxon-Mann-Whitney rank sum test \citep{Wilcoxon1945,Mann1947}.  A random estimator will rank the quiescent galaxy higher half of the time, for a $\Sigma\textrm{ROC}$ of 0.5.

\begin{figure*}[ht!]
\plotone{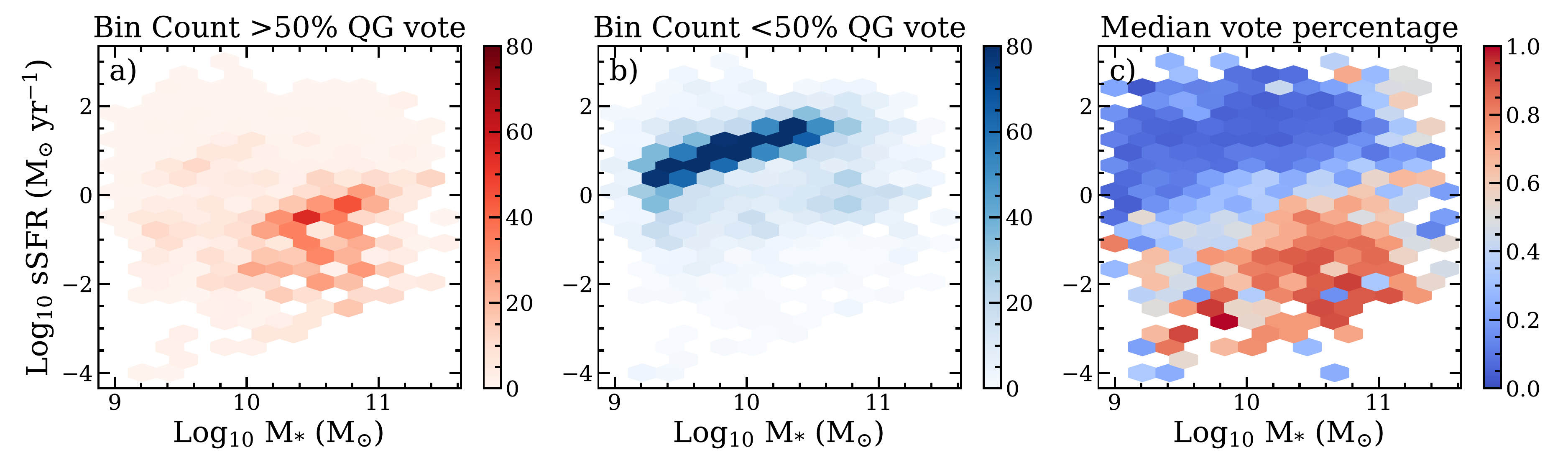}
\caption{(a-b) Distribution of stellar mass and star formation rate for galaxies selected as quiescent or non-quiescent using the t-SNE threshold indicated in Fig. \ref{fig:roc11}a, with shading indicating the number of objects selected.  (c) Median quiescence score using t-SNE selection for the test sample, with a score of 1.0 indicating the highest certainty that an object is quiescent.  Objects with the most uncertain classification lie either near the boundary between star-forming and quiescent, as well as at low mass, where fainter galaxies have higher measurement uncertainties.}
\label{fig:mssel}
\end{figure*}

The t-SNE $\Sigma\textrm{ROC}$ is 0.951 and the photometric template fitting $\Sigma\textrm{ROC}$ selecting quiescent galaxies as those with a low best-fit sSFR is 0.950.  Both are substantially better than random, and t-SNE outperforms sSFR in 63\% of trials from random draws of training and test samples.  The objects for which the t-SNE estimator produces an incorrect classification, as might be expected, are primarily those in two categories: (1) galaxies with high measurement uncertainties, typically fainter and thus lower-mass galaxies; and (2) galaxies near the boundary between star-forming and quiescent, for which even small differences in SFR would change their classification {(Fig. \ref{fig:mssel})}. 

In Fig. \ref{fig:roc11}, we have attempted to optimize the choice of hyperparameters for t-SNE to produce a clean separation.  With a different choice of hyperparameters, the quality of the estimator would be decreased, much as it would be if inferior templates were used for photometric template fitting.  The ideal choice of hyperparameters depends upon the sample, but can be estimated using the following heuristics:
\begin{itemize}
    \item The perplexity should represent the number of neighbors that should be considered informative as to the average galaxy.  Thus, if a survey is doubled in area with otherwise identical detection properties, the optimal perplexity will also approximately double.
    \item The neighborhood radius for determining which objects are sufficiently close on t-SNE map to vote on quiescence should be chosen so the the typical number of objects within that radius matches the perplexity.
    \item The threshold can then be set to different values depending on the relative importance of completeness and quality.  A threshold set near the total fraction of quiescent galaxies, or equivalently the voting score that would result if neighbors are chosen randomly from the entire sample rather than from the t-SNE map, will typically maximize the sum of TPR and TNR.
\end{itemize}

A reasonable interpretation is that the training sample used by t-SNE has comparable, but slightly more, information about quiescent galaxies than the models used to produce templates. Depending upon which regime is most useful, that information can be used to make either slightly better high-quality or high-completeness samples, but t-SNE must be tuned for that specific purpose.   

\subsection{Comparison of Estimators in a Novel Domain}
\label{subsec:test12}

We now consider these estimators in a domain which is primarily unexplored.  Both the training and test samples are drawn from the same rest-frame ULTRAVISTA catalog, but the training sample is drawn at a photometric redshift of $0.9 < z_\textrm{phot} < 1.1$ and the test sample is drawn at $1.9 < z_\textrm{phot} < 2.1$.  This test is designed to explore the utility of these estimators in finding quiescent galaxies in a new, higher-redshift regime, on the basis of what has been learned about them at lower redshifts.  As a result, although hyperparameters of the t-SNE map were carefully chosen in order to optimize the lower-redshift quiescent galaxy selection, we have then frozen them rather than selecting new hyperparameters for the $z \sim 2$ test in order to provide a fair test of a truly unexplored domain in which there is no training sample to calibrate against.

All estimators considered are predicated on the idea that high-redshift quiescent galaxies look sufficiently similar to low-redshift counterparts that it will be possible to recognize them without high-redshift examples.  For photometric template fitting, the assumption is that quiescent galaxies at different redshifts might possibly have dissimilar properties apart from their low star-formation rates, but that they will be driven by similar astrophysics.  Therefore, the same stellar population synthesis codes, extinction laws, etc. can be used to produce valid templates.  The other two estimators make no direct assumptions about the underlying astrophysics and instead assume that high-redshift quiescent galaxies will have similar SEDs, in a holistic way for t-SNE and in specific bands for UVJ selection.

\begin{figure*}[ht!]
\plotone{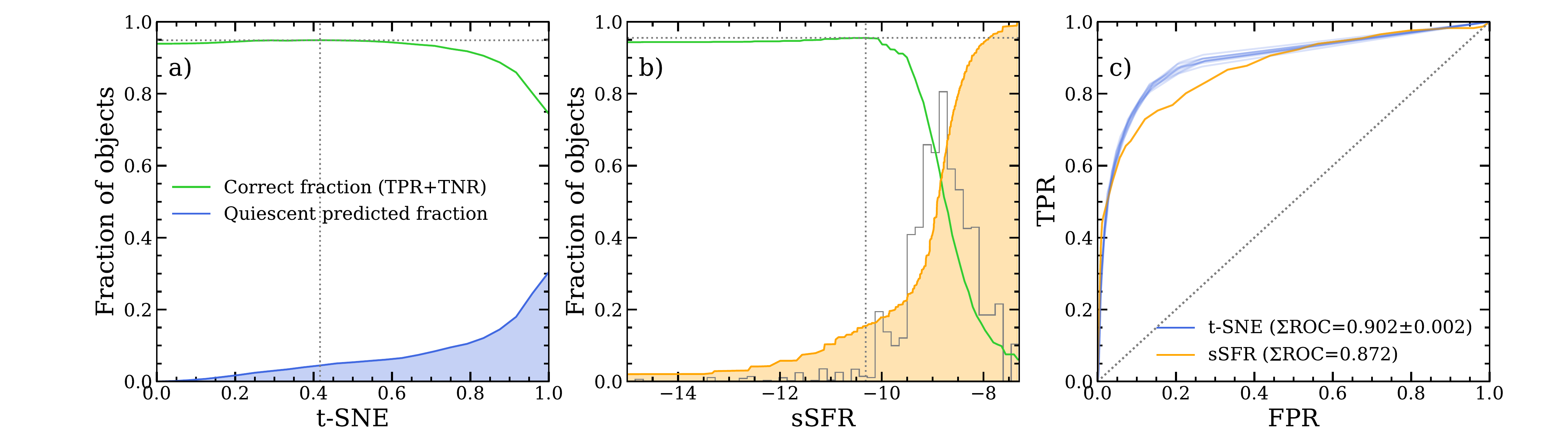}
\caption{(a-b) Correct (sum of of true positive and true negative) prediction fractions for (a) t-SNE and (b) photometric template fitting (green) as a function of the threshold used for a training sample drawn from the $0.9 < z <1.1$ ULTRAVISTA rest-frame catalog and test sample drawn at $1.9 < z <2.1$ from the same catalog.  The sSFR distribution (black, panel b) and cumulative distribution (blue/orange in panels a/b) are also shown, as well as the maximum TPR+TNR achieved and their corresponding thresholds (grey dotted lines). (c) Receiver operating characteristic (ROC) curve for both methods.  t-SNE (blue) is dominant over photometric template fitting (orange) for any choice of optimal true positive rate, with different choices of initial conditions for t-SNE algorithm having negligible impact on the ROC curve.} \label{fig:roc12}
\end{figure*}

As in \S~\ref{subsec:test11}, photometric template fitting and t-SNE both require a choice of threshold and can be assessed through analyzing the true and false positive rates as a function of threshold (Figure \ref{fig:roc12}).  The optimal threshold is lower here, because the overall fraction of true quiescent galaxies is lower in the sample.  In general, the optimal threshold will lie close to the point at which neighbors are consistent with being randomly drawn from the full sample, including both quiescent and non-quiescent galaxies.  This also means that the optimal threshold will depend not just on redshift, but also the detection limit.  Even in a novel domain, some prior expectation about the fraction of quiescent galaxies is required for optimal t-SNE selection.

The resulting ROC curve indicates that t-SNE is a dominant selection mechanism, and allows substantially larger high-quality samples.  
For this unexplored domain, the t-SNE $\Sigma\textrm{ROC}$  is 0.915 and the template fitting $\Sigma\textrm{ROC}$ is 0.871.  In this case, t-SNE even with fixed parameters is dominant over sSFR selection; for any choice of ideal true positive rate, t-SNE selection will have a lower false positive rate than photometric template fitting (Fig. \ref{fig:roc11}c).  The information advantage in favor of color space rather than model space is now sufficiently large that the t-SNE hyperparameters no longer need to be tuned for a specific use case in order to substantially outperform template fitting.

Both values of $\Sigma\textrm{ROC}$ are substantially lower than in the well-sampled domain explored at \S~\ref{subsec:test11}.  This is due to a combination of several factors.  For both methods, $z \sim 2$ galaxies are generally fainter, and therefore more poorly measured than at $z \sim 1$.  Additionally, both methods in different ways assume that high-redshift galaxies look like those at low redshift.  For t-SNE, the assumption is indeed that they are identical in (rest-frame) color space.  For template fitting, this same assumption is instead expressed in the choice of templates, with the assumption that physical models developed using a combination of theory and observed spectra of more local quiescent galaxies continue to describe those at higher redshift.  Finding that both methods are broadly successful at $z \sim 2$ confirms that the colors of quiescent galaxies perhaps change slightly, but do not change substantially between $z =1$ and $z = 2$.  

\subsection{Improvement over Two-Color Selection}
\label{subsec:testuvj}

For completeness, it is also important to show that t-SNE indeed provides an improvement over color-color selection.  It should be expected that t-SNE will perform better than, e.g., UVJ selection because t-SNE is using more information, and is allowed to construct its quiescent locus by combining many small regions of quiescent galaxies rather than one, continuous region.  Thus, it should be hoped that t-SNE will be able to remove many of the UVJ-selected objects with MIPS detections.

However, it is perhaps not obvious {\em a priori} that these advantages must provide a significant improvement.  The colors of quiescent galaxies are dominated by very old stellar populations, and the color of an aging stellar population changes quickly for young populations but is nearly constant for very old ones.  As a result, the observed SEDs of quiescent galaxies look very similar to each other, and predicting the full SED of a quiescent galaxy from a small number of bands is easier than doing so for a star-forming galaxy.  Thus, the additional bands might be mostly redundant information.  If the quiescent locus is tight, there may be negligible benefit to instead describing it as the union of many small neighborhoods and omitting galaxies that lie in between.  Thus, it is necessary to confirm that t-SNE selection truly outperforms UVJ.

Since the quiescent galaxy training sample already includes a two-color (UVJ) selection, t-SNE is trained in part with the goal of reproducing that selection.  Indeed, if UVJ selection were the only criterion used, t-SNE would reproduce it almost exactly, since UVJ selection also corresponds to a region of the full color space described by all bands.  Instead, we test whether t-SNE provides an improvement over UVJ by examining the interlopers with MIPS detections indicating hot dust and therefore that the object is not a quiescent galaxy.  

\begin{figure*}[ht!]
\plotone{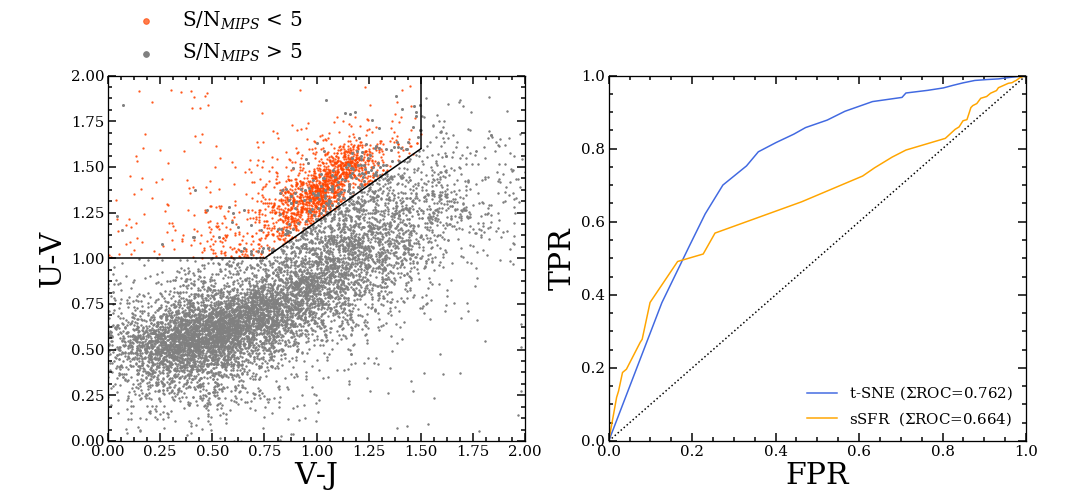}
\caption{(a) UVJ diagram for ULTRAVISTA galaxies at $z\sim 1$.  Galaxies with MIPS detections (gray) are scattered throughout the region and well-mixed with the remainder of the population (red).  (b) Receiver operating characteristic (ROC) curve for t-SNE when selecting non-MIPS detected galaxies from a sample that was previously UVJ-selected.  The t-SNE estimator is still trained on a training sample including all types of galaxies, but the ROC curve is only calculated from UVJ-selected galaxies.  Given a random pair of UVJ-selected galaxies, one with a MIPS detection and one with no MIPS detection, t-SNE will identify which one has the MIPS detection 76.2\% of the time.  Template fitting also provides an improvement, but a smaller one, correctly identifying the MIPS detection for 66.4\% of possible pairs.  For comparison, UVJ alone makes no distinction between objects which pass its selection, so it would correctly identify the MIPS detection exactly 50\% of the time and would correspond to the dotted line.}
\label{fig:uvjrocc}
\end{figure*}

For UVJ selection, these interlopers are well mixed with true quiescent galaxies, so that it would not be possible to produce a high-quality sample with a more restrictive cut in UVJ (Fig. \ref{fig:uvjrocc}a).  However, t-SNE is able to find some regions which do have lower interloper densities, so that it can produce a higher-quality sample (Fig. \ref{fig:uvjrocc}b).  

Selecting quiescent galaxies from a catalog of only UVJ-selected galaxies is a much more difficult problem than selecting them from the full catalog.  The full catalog contains many galaxies which can be very easily rejected as quiescent candidates, whereas the interlopers which pass UVJ selection look far more similar to true quiescent galaxies.  Thus, t-SNE is less effective at this separation, with a $\Sigma\textrm{ROC}$ of 0.762, compared with $\Sigma\textrm{ROC}=0.952$ for the full catalog.  However, this is still a significant improvement upon UVJ selection; it means that given the (rest-frame) optical photometry for a true quiescent galaxy and a dusty star-forming galaxy which also passed UVJ selection and asked to select which one is truly quiescent, t-SNE will make the correct selection 76.2\% of the time.  Template fitting also provides an improvement over UVJ selection, but a smaller one, with a $\Sigma\textrm{ROC}$ of 0.664.  Both are an improvement upon UVJ selection alone, which makes no distinction between objects which pass its selection and thus would correctly identify the MIPS detection exactly 50\% of the time.

Clearly the best estimator of whether MIPS will detect an object is based on MIPS observations, and in practice neither t-SNE nor sSFR estimators would be used instead.  However, coverage is not always available, or is not available at sufficient depth, when determining quiescent candidates from photometry.  Indeed, the rationale behind using template fitting is that it should be possible to predict 24$\mu$m flux from optical and near-infrared photometry.  These results show that t-SNE is a much better estimator of 24$\mu$m flux than photometric template fitting.

\section{Results \& Discussion}

In this work, we develop a new, machine learning-based method for selecting quiescent galaxies from optical photometry.  This method provides an improved catalog over two-color selection by identifying and rejecting many (but not all) of the dusty star-forming galaxies which contaminate two-color samples.  The choice of t-SNE threshold also provides a tradeoff between sample size and sample quality which does not exist in two-color classification.

The efficiency of this t-SNE is compared with photometric template fitting, which similarly allows a tradeoff between quality and quantity, with more restrictive cuts on best-fit sSFR or SFR providing a smaller but higher-quality sample.  We find that in a well-explored domain, in which there is already a large training sample available at the same redshift, t-SNE outperforms photometric template fitting in 63\% of trials.  In a novel domain, using $z\sim 1$ galaxies as a training sample to select $z \sim 2$ quiescent candidates, t-SNE is dominant over template fitting, in that for any choice of sample size, t-SNE will produce a sample with fewer interlopers. 

It should also be stressed that t-SNE was not evaluated here under optimal conditions.  The mapping was based upon a limited number of bands, with IR bands available to template fitting excluded from the t-SNE mapping (since t-SNE was asked in part to make a prediction of 24$\mu$m luminosity), whereas other surveys often provide more information.  Further, because t-SNE needs to compare objects on a similar vector basis, it was necessary to provide it with rest-frame optical measurements.  For objects with catastrophic errors in photometric redshifts, t-SNE was simply provided incorrect data, so that some of the objects for which template fitting fails automatically failed t-SNE as well.  Still, the result is that t-SNE provides an improvement over both color selection and photometric template fitting under essentially all conditions for which it was tested.

An additional issue for both template fitting and t-SNE is computational complexity.  Two-color selection is very simple (although in practice, two-color selection is done in the rest frame, and may require template fitting to be performed prior to selection), and adding objects to an existing sample is also quick.  Photometric template fitting is a slow process, with the quality of the fit often determined by the limited set of templates that can be considered given available computing time.  However, adding one new object to an existing catalog only requires running template fitting on that one object, which can typically be done in minutes at reasonable quality.  
Because t-SNE does not produce a static map, the addition of even one new object requires reoptimizing the entire map, which is the time-consuming step.  Formally, t-SNE runs in $\mathcal{O}(n^2)$, and the runtime is already prohibitive for the $\sim 10^4$ objects in the samples shown here.  Approximations (e.g., Barnes-Hut; \citealt{Barnes1986}) exist to produce reductions to $d \leq 3$ dimensions in $\mathcal{O}(n\log n)$ and were used here to produce a runtime of a few minutes.  However, this would still be an issue for the entire $> 10^6$ object COSMOS catalog or upcoming catalogs from LSST, Euclid, etc.  For such surveys, an alternative algorithm such as the self-organizing map \citep{Masters2015} which produces a static mapping is likely a better choice if real-time decision-making is required for an individual object.  

\subsection{Where Does the Improvement Come From?}

Understanding this improvement requires evaluating how much information is being used and how useful that information is for quiescent selection.  The improvement in t-SNE over color selection is straightforward, since t-SNE is using all of the available information instead of only some bands and doing so in a more flexible manner.  The ability of t-SNE to do this indicates that the three assumptions in \S~\ref{sec:estimators} are generally true, which should not be surprising because these same assumptions are required for photometry to be capable of building a catalog.

Photometric template fitting and t-SNE are much less similar, since they arise from two different ways of modeling galaxies.  Photometry describes galaxies in physical parameter space, using astrophysical modeling to transform those parameters into colors on the basis of the knowledge that human astronomers have built up about galaxy evolution.  t-SNE models galaxies purely empirically in color space, with no astrophysical knowledge used at any stage of the process, so that the information depends purely on the sample size.  

The greater success of t-SNE means that at present, our astrophysical models provide less information about quiescent galaxies and dusty star-forming interlopers than photometric catalogs, with the gap increasing towards high redshift.  This is the opposite of the current situation for redshift determination, for which astrophysical models provide more information and templates produce better fits.  Whether this remains true going forward will depend upon the rate at which models improve compared with the rate at which catalogs become larger and higher quality. 

On the other hand, it should be noted that the performance of the two-color selection can be improved without resorting to machine learning. In case of a rich multi-wavelength  baseline as in COSMOS, the  astrophysical model limitation  mentioned above can be minimized by estimating rest-frame magnitudes from the nearest observer's frame band \citep[see  e.g.][]{Davidzon2017}. Moreover, defining the quiescent \textit{locus} in the $NUV-r$ vs.\ $r-J$ diagram (NUVrJ) instead  of UVJ dramatically reduces the contamination fraction, since $NUV-r$ probe shorter star formation time scales and it is more sensitive to fast quenching processes  \citep[see discussion in][]{Moutard16b}. To show that, we compare the NUVrJ catalog of  quiescent galaxies provided by \citet{Laigle2016} with a fiducial t-SNE selection  resulting from a  threshold equal to 0.46, which maximizes TPR+TNR at $0.9<z<1.1$.  For the same validation sample used above (1,685  quiescent galaxies) the two methods have a similar fraction of interlopers ($\sim$22\% in both cases) and a comparable level of completeness: NUVrJ recovers 84\% of the ``true'' quiescent galaxies while t-SNE  (in the 0.46 threshold configuration) about 79\%. 

It is perhaps more surprising that training t-SNE at $z \sim 1$ and predicting whether $z \sim 2$ galaxies are quiescent still outperforms template fitting. This means that at least in (rest-frame) color space, galaxies at $z \sim 2$ are nearly identical to those at $z \sim 1$.  If this were to continue to hold at much higher redshift, it means that t-SNE would be a good method for selecting high-redshift quiescent candidates from rest-frame optical data using, e.g., the James Webb Space Telescope.  However, one should use considerable caution here: because it has no astrophysical knowledge, t-SNE is only capable of selecting high-redshift quiescent galaxies {\em which look like low-redshift examples}.  If high-redshift quiescent galaxies have different astrophysical properties and therefore exhibit different colors, they will not be selected by t-SNE, but could still be selected by template fitting if these new properties are well described by models.  Similarly, if the low-redshift training sample definition of ``true'' quiescent galaxies is flawed or incomplete, then t-SNE will attempt to faithfully reproduce that flawed selection at high redshift.

\subsection{Additional Considerations}

It should be noted that, as suggested by Fig. \ref{fig:properties}, t-SNE may be used to estimate many galaxy properties apart from quiescence.  A full discussion of these possibilities is beyond the scope of this paper.  However, as one example, t-SNE is used to estimate MIPS detection alone (for both quiescent and star-forming galaxies). In this way we can evaluate the method against a quantity that has been measured directly from telescope images. 

\begin{figure}
\plotone{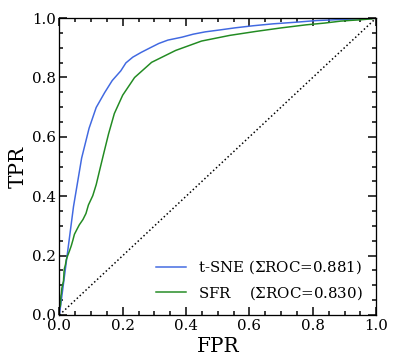}
\caption{Receiver operating characteristic (ROC) curve for t-SNE selection and photometric template fitting's best-fit SFR for selecting MIPS detected galaxies from the full $0.9 < z < 1.1$ sample.}   
\label{fig:mipsrocc}
\end{figure}

For MIPS detections as well, t-SNE is able to produce a meaningful predictor (Fig. \ref{fig:mipsrocc}).  Since a MIPS detection corresponds to a high SFR, not sSFR, it is compared against the set of galaxies with best-fit SFR above some threshold using photometric template fitting.  The comparison has a simple qualitative purpose: the best-fit SFR is by construction a poorer predictor of 24 $\mu$m emission, as the latter depends also on the amount of dust (which span a large range in galaxies at a given SFR).   The t-SNE $\Sigma\textrm{ROC}$ is 0.881 and the template fitting $\Sigma\textrm{ROC}$ is 0.830.  

In any regime in which target galaxies look similar to an existing catalog of examples, dimensionality reduction provides an alternative selection mechanism and alternative method for determining physical properties.  Because quiescent galaxies are all characterized by similar (very old) stellar populations with little AGN contamination, their selection presents an ideal use case for these new methods, and at this point dimensionality reduction provides superior classification to existing techniques.

It should be noted that it is not possible to provide a simple prescription of applying  dimensionality reduction and subsequent selection to a new catalog.  Rather, doing so effectively requires carefully tuning t-SNE hyperparameters to match the expected properties of that catalog.  For example, perplexity needs to be tuned in order to ensure that the number of objects strongly influencing the locations on the t-SNE mapping matches the expected number of meaningful neighbors.  Similarly, the choice of t-SNE threshold depends upon the expected fraction of true quiescent galaxies in the sample.  Properly applying t-SNE to, e.g., CANDELS \citep{Koekemoer2011,Grogin2011} will almost certainly yield an improved estimator, but additional optimization will be required to produce that estimator.

Finally, it should be stressed that other techniques within the family of machine learning methods hold the possibility of substantial further improvement.  It can be more difficult to understand where the improvement is coming from using t-SNE, but an initial exploration (beyond the scope of this current paper) suggests that it may be possible to produce a further improved estimator.  The best methods for selecting quiescent galaxies in poorly explored domains such as at high redshift, contrary to conventional wisdom, might not rely on improved model making or on expensive observations of a few specimens.  Instead, future photometric surveys will probe those domains with enough statistics so that the galaxy color space, albeit unclassified, might be analyzed by means of t-SNE or other manifold learning algorithms to identify galaxy classes with no need for templates.

\vspace{12pt} 
The authors wish to thank Johann Bock Severin, Gabe Brammer, Beryl Hovis-Afflerbach, Adam Jermyn, Christian Kragh Jespersen, Vasily Kokorev, Allison Man, Georgios Magdis, and Jonas Vinther for useful discussions.  CLS and ST are supported by ERC grant 648179 "ConTExt".  The Cosmic Dawn Center (DAWN) is funded by the Danish National Research Foundation under grant No. 140.  JM is supported by the Jonathan Baker Excellence in Physics Fund.

\bibliographystyle{apj}
\bibliography{ref}



\end{document}